\documentclass[twocolumn,showpacs,preprintnumbers,amsmath,amssymb,apsrev]{revtex4}

\usepackage{graphicx}
\usepackage{dcolumn}
\usepackage{bm}

\begin{document}

\preprint{}

\title{Breakdown of Energy Equipartition in a 2D Binary Vibrated Granular Gas}

\author{Klebert Feitosa}
 \email{kfeitosa@physics.umass.edu}
\author{Narayanan Menon}%
 \email{menon@physics.umass.edu}
\affiliation{%
Department of Physics, University of Massachusetts, Amherst, Massachusetts 01003-3720
}%

\date{November 20, 2001}

\begin{abstract}
We report experiments on the equipartition of kinetic energy between grains made 
of two different materials in a mixture of grains vibrated in 2 dimensions.  In 
general, the two types of grains do not attain the same granular temperature, $T_{g}
 = 1/2 m<\bf{v}^2>$.  However, the ratio of the two temperatures is constant in the bulk 
of the system and independent of the vibration velocity.  The ratio depends 
strongly on the ratio of mass densities of the grains, but is not sensitive to the 
inelasticity of grains.  Also, this ratio is insensitive to 
compositional variables of the mixture such as the number fraction of each 
component and the total number density.  We conclude that a single granular 
temperature, as traditionally defined, does not characterize a multi-component 
mixture.
\end{abstract}


\maketitle

When two different gases are put in thermal contact with each other they 
eventually reach thermal equilibrium, with equipartition of kinetic energy 
between the gas molecules and a single temperature for the system as mandated by
 the zeroth law of thermodynamics. In this article we present an experiment 
designed to test the validity of energy equipartition in a mixture of two kinds 
of macroscopic grains driven into a non-equilibrium steady state by external 
vibration. Kinetic theories for granular systems have long used the idea of a
 granular temperature defined as the average kinetic energy of the particles, 
$T_{g}=1/2m<\bf{v}^2>$; the value of $T_{g}$ in a given system is determined by a balance 
between the power input by external driving sources and the energy lost in 
interparticle collisions. The results of such calculations of transport in single-component 
systems have been validated in both simulations \cite{HelalBiben97,BizonShattuck99-1} and
 experiments \cite{RouyerMenon01RerichaBizon01YangHuan01}. It is natural to ask 
whether the variable $T_{g}$ obeys a zeroth law, and in particular, whether a mixture
 of grains is characterized by a single, shared, granular temperature.

The earliest theoretical developments of kinetic theories for binary mixtures of 
spheres \cite{FarrellLun86,JenkinsMancini89} or of discs \cite{JenkinsMancini87}, 
start with a clearly identified assumption that a single temperature variable, $T_{g}$, 
characterizes the entire mixture. Most subsequent work preserves this 
assumption. An important exception is the work of Garzo and Dufty \cite{GarzoDufty99} who study 
theoretically the ``cooling'' of a binary granular mixture from an 
initial distribution of particle velocities. They conclude that the cooling rate 
of each species in the mixture is the same, although their temperatures are 
different throughout the cooling process. Experiments and simulations on binary 
mixtures of granular materials have primarily focused on segregation and mixing 
of different types of grains \cite{OttinoKhakhar00}, and for the most part, have not considered the issue 
of equipartition.  However, an experiment on a vibrated monolayer of grains by Losert et al. 
\cite{LosertCooper99} found that grains of two different sizes and masses do not attain
 the same value of $T_{g}$.  In this article, we show that such differences persist 
in the bulk of a 2D system and that in general, the two components of a mixture 
arrive at a ratio of granular temperatures, $\gamma \neq 1$. We explore the dependence of the 
temperature ratio, $\gamma$,  on particle properties as well as on compositional parameters 
of the mixture. Our data argue for a redefinition of the granular 
temperature in order to accommodate extension to a multi-component system.


\begin{table}[b] 
\caption{\label{table1} Some material properties of the spheres used in the experiment.}
\begin{ruledtabular}
\begin{tabular}{lrcc}
Particle& Mass & Effective 
& Mass ratio \\
~&[$mg$]~&inelasticity \footnote{The effective inelasticity quoted is the fraction of energy lost in a collision in
 single component system, averaged over all collisions. Since relative velocity of the spheres, the impact parameter,
 and rotational motion all affect the inelasticity of a given collision, the number quoted above is the mean of a 
 wide distribution.}
& w/ glass\\
\hline
Glass& 5.24 & 0.17 &-\\
Alum& 5.80 & 0.31 & 0.92\\
Steel& 15.80 & 0.21 & 0.33\\
Brass& 18.00 & 0.39 & 0.28\\
\end{tabular}
\end{ruledtabular}
\end{table}


We make binary mixtures of spherical glass balls with aluminum, steel or brass 
balls. The two relevant material properties being varied are mass and 
inelasticity (see Table \ref{table1}), and the two compositional parameters being varied 
are the number fraction of each component, $x$, and the number density, 
specified in terms of the the average occupied area fraction, $\rho_{avg}$, of the particles in the cell.  To 
determine the importance of particle material properties we hold average number 
density and number fraction fixed and compare mixtures of steel (heavy-elastic) and 
glass (light-elastic), brass (heavy-inelastic) and glass (light-elastic), and 
aluminum (light-inelastic) and glass (light-elastic) in terms of the $T_{g}$ of each 
component and the temperature ratio $\gamma$. Next, to determine whether failure of 
equipartition is a bulk effect, we hold the number fraction fixed for a steel-glass 
mixture and measure the temperature ratio as the number density of balls is varied.
Finally, to test the dependence of the temperature ratio on the relative fraction 
concentration, we hold the number density fixed for a brass-glass pair, 
and measured the temperature ratio of the components as 
the number fraction of brass is varied. 

\begin{figure}
\includegraphics[width=.5\textwidth]{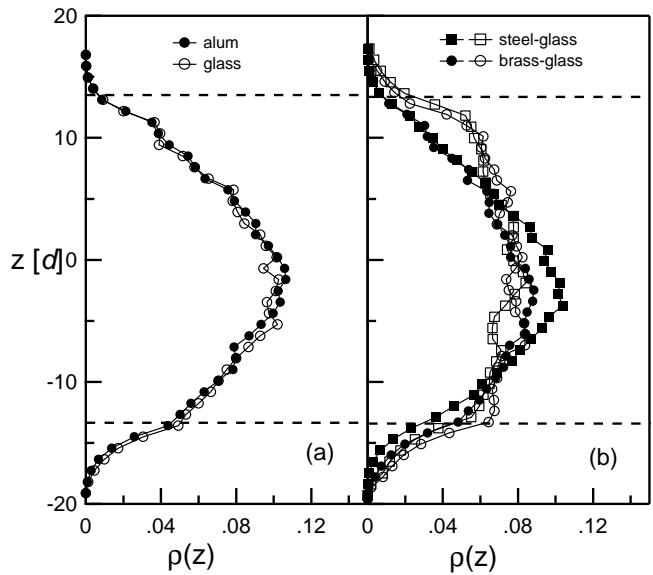}
\caption{\label{density} (a) Density profile, $\rho$, of alum-glass ($\rho_{avg}=0.096$) versus height, z (in $d$).
 (b) Density profile, $\rho$, of steel-glass ($\rho_{avg}=0.096$), and brass-glass
($\rho_{avg}=0.049$, scaled) versus height, z. Dashed lines represent the limits 
of the vibrating walls. $\Gamma = 56~g$ and $x=1/2$ for all 3 pairs.}
\end{figure}


 The balls (diameter: $d=1.600 \pm 0.002~mm$) are confined to a vertical, 
rectangular, Delrin cage (32 $d$ high x 48 $d$ wide x 1.1 $d$ thick) sandwiched 
between two parallel plates of anti-static coated Plexiglas. An electromechanical 
shaker (LDS 500L) vibrates the cage vertically at a frequency of $60~Hz$ and 
amplitudes up to $2.4~d$, producing maximum accelerations, $\Gamma$, and velocities, $v_{0}$, 
of $56~g$ and $1.45~m/s$ respectively. The motions of the balls are recorded with a 
monochrome high-speed camera (Kodak Motioncorder) at a rate of 2000 frames/s. 
Each ball is located with a precision of 0.03 $d$ ($\approx 40~\mu m$). Glass balls are 
unambiguously distinguished from their metallic partners by using a combination 
of reflected and transmitted light, and by thresholding appropriately. Most of 
the results we discuss here are taken in a rectangular (10~$d$ x 20~$d$) window 
centered in the geometrical center of the cell.

\begin{figure}
\includegraphics[width=.5\textwidth]{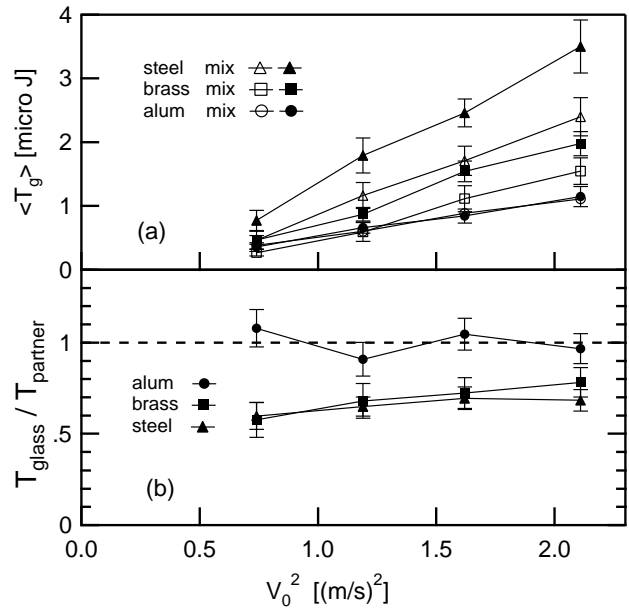}
\caption{\label{temp-ratios} (a) Average granular temperature $<T_{g}>$ versus squared vibration velocity 
$v_{0}^2$ for alum-glass, brass-glass, and steel-glass mixtures. Open symbols and 
filled symbols represent $T_{g}$ for glass and the metal partner, respectively. 
$\rho_{avg}=0.096$ and $x=1/2$ for all mixtures. (b) Ratio between glass temperature 
and partner temperature, $\gamma= T_{glass}/T_{partner}$, versus the squared vibration velocity, $v_{0}^2$.}
\end{figure}

The data presented in this article are all taken at relatively large values of 
excitation, corresponding to $\Gamma > 32~g$ and $v_{0} > 0.86~m/s$.  Beyond this scale of 
excitation, the vertical density profile does not vary much and 
tends to an asymptotic profile that is symmetric about the center of the cell, 
as shown in Fig. \ref{density}. The density profile is identical for two different 
species when their masses are 
matched, even if they have different inelasticity, as in the case of aluminum-glass mixture 
shown in Fig.\ \ref{density}(a). In contrast, the profiles of 
the other two mixtures (steel-glass and brass-glass) in Fig.\ \ref{density}(b), show a slightly higher 
concentration of heavy particles in the center of the cell and a more uniform 
distribution of light particles throughout the cell. However, in all three 
mixtures, while there are differences in the overall density distribution, the 
particles in the mixture are locally well-mixed, with no obvious tendency for 
balls of a particular type to cluster \footnote{By  contrast, at low accelerations, there is a clear 
tendency toward complete segregation, as in the simulations of S. McNamara and S. Luding,
\textit{IUTAM symposium on Segregation in Granular Flows}, eds. A. D. Rosato and D. L.
Blackmore, (Kluwer Academic Publishers, Dordrecht, 2000), p. 305; S. Luding \textit{et al.}, \textit{ibid.} p. 297.} 
when vigorously excited as in this case.  Finally, we have not observed any 
gradients of velocity or density in the horizontal direction.

The principal observation of our experiment is that the two components of a 
binary mixture do not always equilibrate to the same granular temperature. 
Figure \ref{temp-ratios}(a) shows $T_{g}$ of three pairs of materials as a function of the squared 
vibration velocity of the cell, $v_{0}^2$, with the number density being held fixed at $\rho_{avg}=0.096$ and
the number fraction of glass balls fixed at $x=1/2$. The 
granular temperature $T_{g}$ is proportional to the squared vibration velocity $v_{0}^2$ with the 
proportionality constant being a function of the inelasticity of the balls and number density 
$\rho_{avg}$. As shown in Fig.\ \ref{temp-ratios}(b), $T_{g}$ equilibrates for the alum-glass mixture, but does not 
equilibrate for the steel-glass and brass-glass mixtures. In terms of particle properties, equipartition is 
achieved when the mass ratio is close to one  ($m_{glass}/m_{alum}=0.92$), and fails
when the masses are mismatched ($m_{glass}/m_{steel}=0.29$, $m_{glass}/m_{brass}=0.33$). Even 
when the temperatures of the two components do not equilibrate, the ratio of the 
temperatures do not vary as a function of $v_{0}^2$ within the error bars, 
as shown in Fig.\ \ref{temp-ratios}(b). Furthermore, the ratio of temperatures are very close to one another 
($\gamma_{g-s}=T_{glass}/T_{steel}= 0.66 \pm 0.06$ and $\gamma_{g-b}=T_{glass}/T_{brass}=0.69 \pm 0.09$), 
even though the inelasticity of the metal partner in each case is quite 
different (Table \ref{table1}). We conclude that the temperature ratio differs from one 
when the mass ratio departs significantly from one, but does not depend 
sensitively on the relative value of the inelasticity of the two components.

\begin{figure}[t]
\includegraphics[width=.5\textwidth]{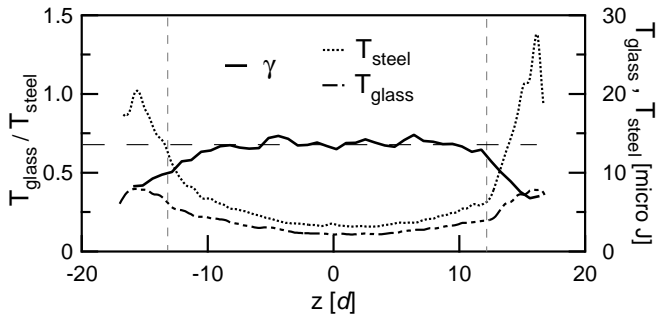}
\caption{\label{temp-cell} Vertical temperature profiles for a steel-glass mixture. 
Dotted and dot-dash lines represent $T_{steel}$ and $T_{glass}$ and the full line represents 
the ratio between the glass and steel temperatures. Dashed vertical lines 
represent the limits of the vibrating walls. For this system $\rho_{avg}=0.096$, $x=1/2$ and $\Gamma=56~g$.}
\end{figure}

The failure of equipartition is not merely a boundary effect, nor is the value 
of $\gamma$ that we report in Fig.\ \ref{temp-ratios}(b), an inhomogeneous average over different regions 
of the cell. This is shown in Fig.\ \ref{temp-cell} where the temperature profiles of steel and 
glass are plotted as a function of the height of the cell, z[$d$], for $v_{0}=1.45~m/s$. 
In a collision with a moving wall which is effectively infinitely massive, both species pick up
the same velocity, which implies that the heavier species (steel) picks up greater kinetic energy. The temperature
ratio $\gamma$ is thus different from one at the walls; this is analogous to the previous 
observations of Losert et al \cite {LosertCooper99}. Remarkably, however, the temperature ratio quickly reaches 
a constant value ($\gamma = 0.69$) in the interior of the cell, even as the $T_{g}$ of each component 
continues to decrease toward the middle of the cell as a result of losses from 
inelastic collisions. Therefore the effect of inelastic collisions is to drive 
the value of $\gamma$ in the interior of the cell toward a value different from one, 
which as we mentioned above, depends on the mass ratio of components, but is 
insensitive to their inelasticity, and to the value of $v_{0}$.

\begin{figure}[b]
\includegraphics[width=.5\textwidth]{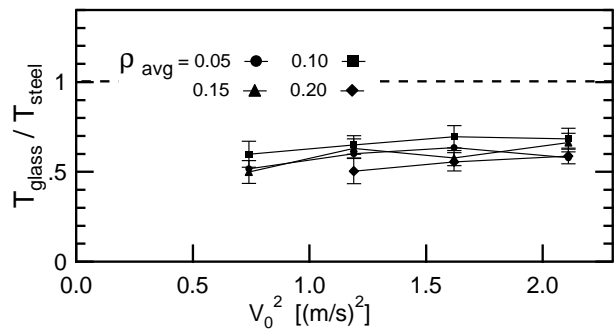}
\caption{\label{temp-steel} Temperature ratio, $\gamma=T_{glass}/T_{steel}$, in a steel-glass mixture with number fraction
 $x=1/2$, plotted against squared vibration velocity, $v_{0}^2$. Different markers 
represent different number densities of the mixture.}
\end{figure}

We pursue further the possibility that the value of $\gamma$ obtained is due to 
boundary effects by increasing the number density, $\rho_{avg}$, of a steel-glass 
mixture holding the number fraction fixed at $x=1/2$. While this is not exactly 
equivalent to increasing the system size, it does tune the ratio of system size 
to mean free path, so that we vary the number of collisions experienced by balls 
in transferring energy from the boundary to the interior. The temperature ratio, 
$\gamma = T_{glass}/T_{steel}$ is plotted against of the square vibration velocity,  $v_{0}^2$, 
in Fig.\ \ref{temp-steel}, for mixtures whose number density, $\rho_{avg}$, varies by a factor of 4. 
Once again, the ratio of the two temperatures are essentially independent of $v_{0}^2$.
Importantly, the 
temperature ratio maintains an average value of $\gamma = 0.60 \pm 0.06$, with no obvious systematic trend 
as $\rho_{avg}$ changes. This result further confirms that the failure of equipartition is 
a bulk effect, and also suggests that the value of $\gamma$ is insensitive to the 
number density of the mixture.

\begin{figure}[t]
\includegraphics[width=.5\textwidth]{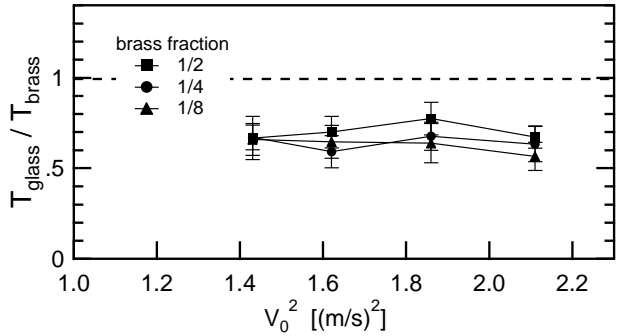}
\caption{\label{temp-brass} Ratio between glass temperature and brass temperature, $\gamma=T_{glass}/T_{brass}$,
versus squared vibration velocity of the cell, $v_{0}^2$. Different markers represent different 
number fractions of brass for the same total number of particles ($\rho_{avg}=0.049$).}
\end{figure}

Finally, we study the effect on equipartition of varying number fraction in a 
mixture at fixed density. Figure \ref{temp-brass} shows the temperature ratio, $\gamma= T_{glass}/T_{brass}$, 
of a brass-glass mixture as a function of $v_{0}^2$, for three different number fractions of 
brass, $x$: 1/2, 1/4, and 1/8. The temperature ratios for the different relative 
concentrations are all the same within experimental errors. This demonstrates 
that the components do not reach thermal equilibrium, even though the relative 
number of collisions with glass balls increase for a given brass ball, as the 
number fraction of brass is decreased. It has been predicted \cite {SantosDufty01}  that 
equipartition does not generally occur even in the tracer limit of a single 
foreign particle in a sea of other particles.  This is consistent with our 
observation that the temperature ratio $\gamma$ does not depend on number fraction. 


As mentioned earlier, Garzo and Dufty \cite {GarzoDufty99} study a freely-cooling
binary system, and find a temperature ratio dependence 
on material parameters as well as on compositional parameters. Since there are no
comparable predictions for a driven system, we have compared our data with these
predictions and find  quite good agreement with our results as a function of material
parameters. However, they predict a strong dependence on number fraction of the 
components, where none is observed. The comparison between theory and experiment 
encourages further theoretical development in driven systems. 

In conclusion, we have observed that energy equipartition does not generally 
hold for a binary vibrated granular system. However, the ratio between the 
steady state values of the granular temperatures of the two components does 
robustly satisfy some general trends. The temperature ratio is constant in the 
interior of a system, and is insensitive to vibration velocity and inelasticity, 
while depending strongly on mass density.  Furthermore, the ratio is insensitive 
to compositional parameters such as number fraction of each component and average 
number density.  This opens the possibility that just as in the work of Garzo 
and Dufty \cite {GarzoDufty99} for a cooling system, there is a well-defined and calculable ratio of 
temperatures achieved in a steady state system.   Thus a modified definition of granular 
temperature with a prefactor 
depending chiefly on material properties, may adequately describe a 
multi-component system.  Numerical simulations could play a useful role, in that 
material properties may be continuously tuned, as opposed to experiments in 
which only a limited pool of materials is readily available.


\begin{acknowledgments}
We gratefully acknowledge support from NSF DMR CAREER 987433, and ACS PRF G33730.  
We also thank J.~L.~Machta for useful conversations and F.~Rouyer for contributions to this work.
\end{acknowledgments}

\newpage 

\end{document}